\documentclass[12pt]{article}
\usepackage{amsmath}
\usepackage{graphicx}
\usepackage{enumerate}
\usepackage{mathtools}
\usepackage[colorinlistoftodos]{todonotes}
\usepackage[a4paper, margin=1in]{geometry}
\usepackage{amssymb}
\usepackage{multirow}
\usepackage{bm}
\usepackage{natbib}
\usepackage{times}
\usepackage{caption}
\usepackage{subcaption}
\usepackage[plain,noend]{algorithm2e}

\title{Adjusted Empirical Likelihood for Time Series Models}

\author{Ramadha D. Piyadi Gamage, Wei Ning \footnote{Corresponding author. Email:
wning@bgsu.edu}\hskip .1in and Arjun K. Gupta\\ Department of
Mathematics and Statistics\\Bowling Green State University, Bowling
Green, OH 43403, USA}

\begin{document}
\maketitle

\begin{abstract}
\noindent Empirical likelihood method has been applied to dependent observations by Monti (1997) through the Whittle's estimation method.
Similar asymptotic distribution of the empirical likelihood ratio statistic for stationary time series has been derived to construct
the confidence regions for the parameters. However, required numerical problem of computing profile empirical likelihood function which involves constrained maximization has no solution sometimes, which leads to the drawbacks of using the original version of the empirical likelihood ratio. In
this paper, we propose an adjusted empirical likelihood ratio statistic to modify the one proposed by Monti so that it guarantees the existence of the solution of the required maximization problem, while maintaining the similar asymptotic properties as Monti obtained. Simulations have
been conducted to illustrate the coverage probabilities obtained by the adjusted version for different time series models which are better than the ones obtained by Monti's version, especially for small sample sizes.
\end{abstract}

\section{Introduction}
\noindent
Empirical Likelihood (EL) method introduced by Owen (1988) has become a widely applicable tool for constructing confidence regions in nonparametric problems due to its appealing asymptotic distribution of the likelihood-ratio-type statistic which is same as the one under the parametric settings. However, the established methods of EL are mainly for independent observations and therefore are difficult to apply to dependent observations such as time series data. Progress in applying the EL method on different time series models has been made by different researchers. For example, Mykland (1995) built up the connection between the dual likelihood and the empirical likelihood through the martingale estimating equations and applied to time series model. Kitamura (1997) proposed the empirical likelihood of the blocks of observations to study the weakly dependent processes. Chan and Ling (2006) derived the empirical likelihood method for \textsc{garch} models. Chan and Liu (2010) considered the Bartlett corrections of empirical likelihood for short memory time series. Monti (1997) extended the EL method to stationary time series by using the Whittle's (1953) estimation method to obtain an M-estimator of the periodogram ordinates of time series models which are asymptotically independent. This method reduced the dependent data problem to an independent data problem. Then the original EL method is applied and similar asymptotic results have been obtained. Chun (2011) extended Monit's results to long-memory time series models. 

However, as Chen et al. (2008) pointed out, computing profile empirical likelihood function which involves constrained maximization requires that the convex hull of the estimating equation must have the zero vector as an interior point. When the required computational problem has no solution, Owen (2001) suggested assigning~$-\infty$~to the log-EL statistic. Chen et al.(2008) mentioned there are two drawbacks in doing so. To remedy the drawback of EL method, Chen et al. (2008) proposed an adjusted empirical likelihood (AEL) for independent observations by adding an artificial term to guarantee the zero vector to be within the convex hull, therefore, the solution always exists. In their work, they have showed the asymptotic results of the AEL are same as that of the EL. Furthermore, this method can achieve the improved coverage probabilities without using Bartlett-correction or bootstrap calibration.

In this paper, we propose an adjusted empirical likelihood method to extend Monti's method by adopting
Chen's idea. The rest of the article is organized as follows. In Section 2, the AEL for M-estimators is derived. Section 3 reviews briefly the Whittle's estimator based on the periodogram. The AEL for a stationary time series model is proposed and the asymptotic distribution of the AEL statistic is established in Section 4. The application to \textsc{arma} models is considered in Section 5. Simulations in Section 6 illustrate the coverage probabilities of the proposed AEL method for invertible moving average (\textsc{ma}) time series model and stationary autoregressive (\textsc{ar}) model with various sample sizes, values of parameters and distributions of the noise term. Comparisons to the EL method, Bartlett-corrected ones have also been made to indicate the advantage of the proposed AEL method. Some discussion is provided in Section 7.

\section{Adjusted Empirical Likelihood For M-Estimators}
\noindent The empirical likelihood (EL) method was introduced by Owen (1988,1990,1991) which combines the reliability from nonparametric methods and flexibility of parametric methods. The most appealing property of the EL method is that the null distribution of the EL ratio statistic follows the standard chi-square distribution under mild conditions similar to the one obtained under the parametric settings. Since then, the EL method has been extensively used in different areas in statistics. See Owen (2001) for more details of the EL method. However, the EL method is designed to deal with the independent observations and has difficulties in applying to dependent observations such as time series. Based on the facts that the periodogram of time series data are asymptotically independent and the parameter of a stationary time series estimated by the Whittle's (1953) method can be viewed as an M-estimator involving periodogram ordinates, Monti (1997) proposed the empirical likelihood version based on M-estimators to reduce a dependent data to an independent data, therefore, Owen's EL method can be applied. Let $ x_1,x_2,...,x_n $ be a set of independent and identically distributed observations from an unknown distribution~$F_0$~and $X\sim F_\beta$ where~$\beta\in \mathbb{B}$~is a~$k-$~vector. $\beta$ is estimated by an $M$-estimator $\beta_n$, which is the solution of
$$\sum_{j=1}^{n}\psi_j(x_j,\beta)=0.$$
With the definition of~$p_j=P(X=x_j),$ the empirical likelihood ratio for any value $\beta \in \mathbb{B}$ is defined by
$$\hat{\lambda}(\beta) = \left. {\sup \prod_{j=1}^{n} p_j} \middle/ {\prod_{j=1}^{n} \frac{1}{n}} \right. = \sup \prod_{j=1}^{n} np_j,$$
subject to the constraints:(i) $\sum_{j=1}^{n}\psi(x_j,\beta) p_j=0$, (ii) $\sum_{j=1}^{n} p_j=1$, and (iii) $p_j \geq 0 $ , ($j=1,2,...n$). The maximization under the Lagrange multiplier method gives
$$ p_j = [n\{1 + \hat{\xi}(\beta)^\prime\psi(x_j,\beta)\}]^{-1} \qquad (j=1,2,...,n), $$
where $\hat{\xi}(\beta)$ is the Lagrangian multiplier satisfying constraint (i). The empirical likelihood ratio statistic is thus defined
$$\hat{W}(\beta) = -2\,\ln {\hat{\lambda}(\beta)} = 2 \, \sum_{j=1}^{n} \ln \{1 + \hat{\xi}(\beta)^\prime\psi(x_j,\beta)\}.$$
Owen (1988) showed that $\hat{W}(\beta)$ is asymptotically distributed as ${\chi}^2_k$. Consequently, an asymptotic 1-$\alpha$ confidence region is given by $\{ \beta \in \mathbb{B} : \hat{W}(\beta) \leq {\chi}^2_{k,1-\alpha} \}$. Under some moment conditions on $\psi_j(x_j,\beta)$ (Owen 2001), the convex hull $\{\psi_j(x_j,\beta), i=1,2,...,n\}$ contains 0 as its interior point with probability 1 as $n \rightarrow \infty$. When the parameter $\beta$ is not close to $\beta_n$, or when $n$ is small, there is a good chance that the solution to constraint (i) doesn't exist which will raise some computational issues as Chen et al. (2008) mentioned. To overcome this difficulty, Chen et al.(2008) proposed an adjusted empirical likelihood (AEL) ratio function by adding $\psi_{n+1}$-th term to guarantee the zero to be an interior point of the convex hull, therefore, the required numerical maximization always has the solution. By doing so, they modified Owen's method and applied to independent observations with the establishment of same asymptotic null distribution of the statistic as Owen obtained. As follows, we adopt their idea to modify Monti's method for dependent observations.

Denote $\psi_j = \psi_j(\beta) =  \psi(x_j,\beta)$ and $\bar{\psi}_n = \bar{\psi}_n(\beta) = \frac{1}{n}\sum_{j=1}^{n}\psi_j $. For some positive constant $a_n$, define
$$\psi_{n+1} = \psi_{n+1}(\beta) =  -\frac{a_n}{n} \sum_{j=1}^{n}\psi_j = -a_n {\bar{\psi}}_n.$$
Here we choose $a_n=\max(1,log(n)/2)$ coupled with a trimmed version of $\bar{\psi}_n$ when appropriate as Chen et al. (2008) suggested. Hence the empirical likelihood ratio for any value $\beta \in \mathbb{B}$ is adjusted to be,
$$\hat{\lambda}(\beta) = \left. {\sup \prod_{j=1}^{n+1} p_j} \middle/ {\prod_{j=1}^{n+1} \frac{1}{n+1}} \right. = \sup \prod_{j=1}^{n+1} (n+1)p_j,$$
where the maximization is subject to:
\begin{enumerate}[(i)]
  \item $\sum_{j=1}^{n+1}\psi(x_j,\beta) p_j=0$,
  \item $\sum_{j=1}^{n+1} p_j=1$,
  \item $p_j \geq 0 $.
\end{enumerate}
Similarly, with the Lagrange multiplier method we obtain
$$ p_j = [(n+1)\{1 + \hat{\xi}(\beta)^\prime\psi(x_j,\beta)\}]^{-1} \qquad j=1,2,...,n+1, $$
where $\hat{\xi}(\beta)$ is the Lagrangian multiplier satisfying,
\begin{equation}
\sum_{j=1}^{n+1}\frac{\psi(x_j,\beta)}{1 + \hat{\xi}(\beta)^\prime\psi(x_j,\beta)}=0.  \tag{2.1}\label{eq:2.1}
\end{equation}
Thus the adjusted empirical likelihood ratio (AEL) statistic is defined by
$$\hat{W}^*(\beta) = 2 \, \sum_{j=1}^{n+1} \ln \{1 + \hat{\xi}(\beta)^\prime\psi(x_j,\beta)\}.$$
where $\psi_{n+1} = -\frac{a_n}{n} \sum_{j=1}^{n}\psi_j = -a_n {\bar{\psi}}_n$ and $a_n=\max(1,log(n)/2)$.

\section{Whittle Likelihood method based on the Periodogram}
\noindent In this section, we just briefly go over the Whittle's estimator based on the periodogram. Let
\begin{equation}
Z_t = \sum_{s=0}^{\infty} \gamma_s a_{t-s}, \qquad (t=...,-1,0,1,...), \tag{3.1}\label{eq:3.1}
\end{equation} be a linear process where $\gamma_0=1,\sum_{s=0}^{\infty} {\gamma}^2_s < \infty$, and $a_t$ is a sequence of independent and identically distributed random variables with $E(a_t) =0,E(a^2_t) = \sigma^2 > 0,E(a^4_t) < \infty$. The spectral density of $Z_t$ is given by
$$g(\omega) = \frac{\sigma^2}{2\pi} {\bigg| \sum_{s=0}^{\infty} \gamma_s \exp(-i\omega s) \bigg|}^2, \qquad \omega\in [ -\pi,\pi].$$
Let $z_1, z_2,...,z_T$ be $T$ observations of the process \eqref{eq:3.1} with sample mean $\bar{z}$. An approximate log-likelihood function is given by (Whittle, 1953)
\begin{equation}
\ln \{L(\beta)\} = -\sum_{j=1}^{n}\ln\{g_j(\beta)\} -\sum_{j=1}^{n}\frac{I(\omega_j)}{g_j(\beta)}, \tag{3.2}\label{eq:3.2}
\end{equation}
where $g_j(\beta)$ is the spectral density defined above and
$$I(\omega_j) = \frac{1}{2\pi T} \bigg[ {\bigg\{\sum_{t=1}^{T}(z_t-\bar{z})\sin(\omega_j t)\bigg\} }^2 + {\bigg\{\sum_{t=1}^{T}(z_t-\bar{z})\cos(\omega_j t)\bigg\} }^2 \bigg]$$
is the periodogram ordinate evaluated at Fourier frequency $\omega_j=2\pi j/T$, $j=1,2,...,T-1.$
\\The Whittle's estimator~$\hat{\beta}$~maximizes \eqref{eq:3.2} over $\mathbb{B}$. Hence, in terms of the $\psi$-functions
$$\psi_j\{I(\omega_j),\beta\}=\bigg\{\frac{I(\omega_j)}{g_j(\beta)}-1\bigg\} \frac{\partial\ln\{g_j(\beta)\}}{\partial \beta},$$
the estimator has the interpretation of an M-estimator from asymptotically independent periodogram ordinates.

\section{Adjusted Empirical Likelihood for Time Series Models}
\noindent When applying to time series data, the $\psi$-functions which define Whittle's estimator vary with $j$. Therefore we have the log-likelihood ratio as
\begin{equation}
\hat{W}^*(\beta) = 2 \, \sum_{j=1}^{n+1} \ln \big[ 1 + \hat{\xi}(\beta)^\prime\psi_j(I(\omega_j),\beta)\big], \tag{4.1}\label{eq:4.1}
\end{equation}
where $\hat{\xi}(\beta)$ satisfies,
\begin{equation}
\sum_{j=1}^{n+1} {[1 + \hat{\xi}(\beta)^\prime\psi_j\{I(\omega_j),\beta\}]}^{-1}\psi_j\{I(\omega_j),\beta\}=0,  \tag{4.2}\label{eq:4.2}
\end{equation}
where $\psi_{n+1} = -\frac{a_n}{n} \sum_{j=1}^{n}\psi_j = -a_n {\bar{\psi}}_n$ and $a_n=\max(1,log(n)/2)$. Using arguments similar to Monti's (1997) and Chen et al. (2008), we can show that
\begin{align*}
\hat{W}^*(\beta) \sim {\chi}^2_k,
\end{align*}
asymptotically, where~$k=\dim {\beta}.$~ The proof is provided in Appendix A.

\section{Confidence Regions for the Parameters of \textsc{arma} models}
\noindent Let $Z_t$ be an \textsc{arma}$(p,q)$ process,
$$\phi(B)Z_t=\theta(B)a_t,$$
where $B$ is the backward shift operator ($BZ_t=Z_{t-1}$), with $\theta(B) = 1-\theta_1B - \theta_2B^2 - ... - \theta_qB^q$ being the moving average operator, $\phi(B)=1-\phi_1B - \phi_2B^2 - ... - \phi_p B^p$ being the autoregressive operator and white noise process $a_t \sim N(0,\sigma^2)$. The spectral density of \textsc{arma}$(p,q)$ model is given by
$$g(\omega,\beta)=\frac{\sigma^2}{2\pi}\frac{{|\theta(e^{-i\omega})|}^2}{{|\phi(e^{-i\omega})|}^2}, \qquad \omega\in[-\pi,\pi],$$
where $\beta=(\phi_1,...,\phi_p,\theta_1,...,\theta_q,\sigma^2)^\prime$. The variance, $\sigma^2$ of the noise is usually considered as a nuisance parameter as it has no effect on the main characteristics of the process other than modifying the scale of the process. Hence, we consider the profile spectral likelihood function. Let $\beta=(\beta_{(1)},\sigma^2)$, where $\beta_{(1)}$ is the parameter of interest. Monti (1997) showed that by maximizing \eqref{eq:3.2} with respect to $\sigma^2$, the spectral log-likelihood function becomes
$$\ln\{\hat{L}(\beta_{(1)})\} = -n\ln\bigg\{ n^{-1}\sum_{j=1}^{n}\frac{I(\omega_j)}{g^1_j(\beta_{(1)})}\bigg\} -\sum_{j=1}^{n}\ln\{g^1_j(\beta_{(1)})\} - n $$
where
$$g^1_j(\beta_{(1)}) = \frac{1}{2\pi}\frac{{|\theta(e^{-i\omega})|}^2}{{|\phi(e^{-i\omega})|}^2}.$$
Consequently, the profile empirical likelihood ratio is given by
\begin{equation}
\hat{W}^*(\beta_{(1)}) = 2 \, \sum_{j=1}^{n+1} \ln \big[1 + \hat{\xi}(\beta_{(1)})^\prime{\tilde{\psi}}_j(I(\omega_j),\beta_{(1)})\big], \tag{5.1}\label{eq:5.1}
\end{equation}
where the estimator of $\beta_{(1)}$ is the $M$-estimator corresponding to the $\psi$-function
$${\tilde{\psi}}\{ I(\omega_j),\beta_{(1)}\} =\frac{I(\omega_j)}{g^1_j(\beta_{(1)})}\bigg[ \frac{\partial\ln\{g^1_j(\beta_{(1)})\}}{\partial \beta_{(1)}} - (n+1)^{-1} \sum_{j=1}^{n+1}\frac{\partial\ln\{g^1_j(\beta_{(1)})\}}{\partial \beta_{(1)}} \bigg],  $$
and $\hat{\xi}(\beta_{(1)})$ satisfies,
$$\sum_{j=1}^{n+1} {[1 + \hat{\xi}(\beta_{(1)})^\prime{\tilde{\psi}}_j\{I(\omega_j),\beta_{(1)}]}^{-1}\tilde{\psi}_j\{I(\omega_j),\beta\}=0 $$
and $\tilde{\psi}_{n+1} = -\frac{a_n}{n} \sum_{j=1}^{n}\tilde{\psi}_j$ and $a_n=\max(1,log(n)/2)$.
Using arguments similar to Monti (1997), we show that $\hat{W}^*(\beta_{(1)})$ is still asymptotically distributed as ${\chi}^2_{p+q}$. The proof is provided in Appendix B.

\subsection{Example}
We consider the example in Box \& Jenkins (1976, Ch 6), which was analyzed by Monti (1997). The data set describes a series of 197 chemical process concentration readings. They propose an \textsc{arma} (1,1) model,
$$Z_t = \phi Z_{t-1} + a_t - \theta a_{t-1}.$$
Figure 1 shows the 90\% adjusted empirical likelihood confidence region (solid line) for a parameter $\beta_{(1)} = (\phi,\theta)^\prime$. It is compared with an empirical likelihood confidence region (dashed line). We observe that the confidence contour based on the proposed AEL contains the one obtained based on the unadjusted EL while retaining the data-driven shape.
\begin{figure}[ht]
\centering
\includegraphics[height=10cm, width=10cm]{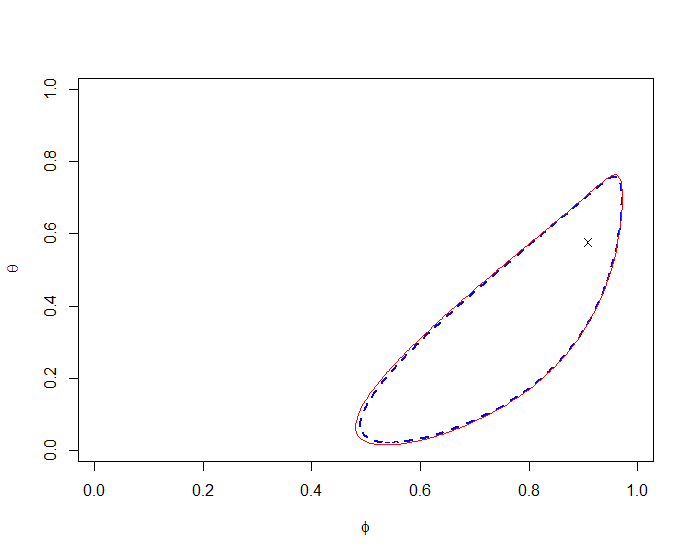}
\captionsetup{justification=centering}
\caption{\small 90\% adjusted empirical likelihood (solid line) and empirical likelihood (dashed line) confidence regions for the parameters of an \textsc{arma(1,1)} model fitted to the series of chemical process concentration readings (Box \& Jenkins, 1976)}
\end{figure}

\subsection{Coverage error of adjusted empirical likelihood confidence intervals}
We also conduct simulations with various values of~$\phi$~and~$\theta$~for \textsc{arma}(1,1) model to illustrate improvement of the AEL over the EL on constructing confidence regions. \textsc{arma} model with $\phi=0.7$ and $\theta=0.5$ is used here. 1000 series of observations are drawn under three different sample sizes: n=40, 70, 100 with errors following the standard normal distribution. $\hat{W}(\beta)$ is calculated at different points over the parameter space $(\phi,\theta) \in \{(0,1) \times(0,1)\}$ and contour plots are produced for each of the three sample sizes using threshold value ${\chi}^2_{2,0.9}$ to construct 90\% confidence region. The empirical likelihood confidence regions are also constructed. Figure 2 shows the 90\% adjusted empirical likelihood confidence regions for sample sizes 40, 70 and 100 under standard normal error distribution. In each case the mean is subtracted from the white noise processes in order to have mean zero for the error terms. Similarly, with the same nominal level, the confidence contours for AEL contain the ones based on the unadjusted EL, especially when the sample size is small.
\begin{figure}[ht]
    \centering
    \begin{subfigure}[b]{0.3\textwidth}
        \includegraphics[width=\textwidth]{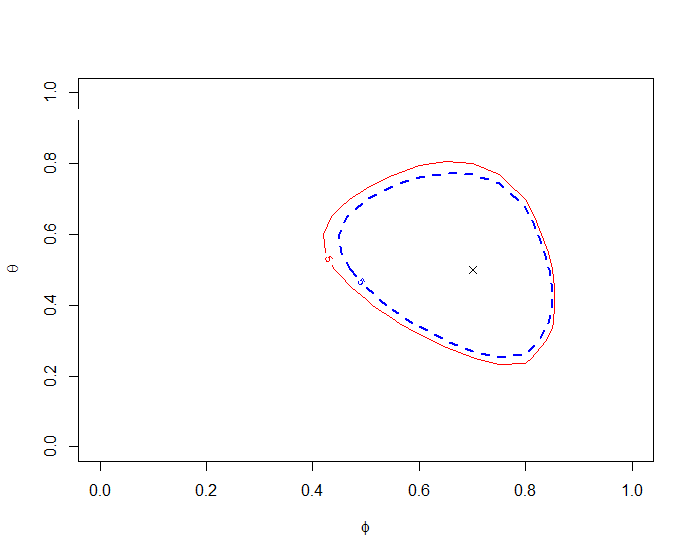}
        \caption{\small n=40}
        \label{fig:n=40}
    \end{subfigure}
    \begin{subfigure}[b]{0.3\textwidth}
        \includegraphics[width=\textwidth]{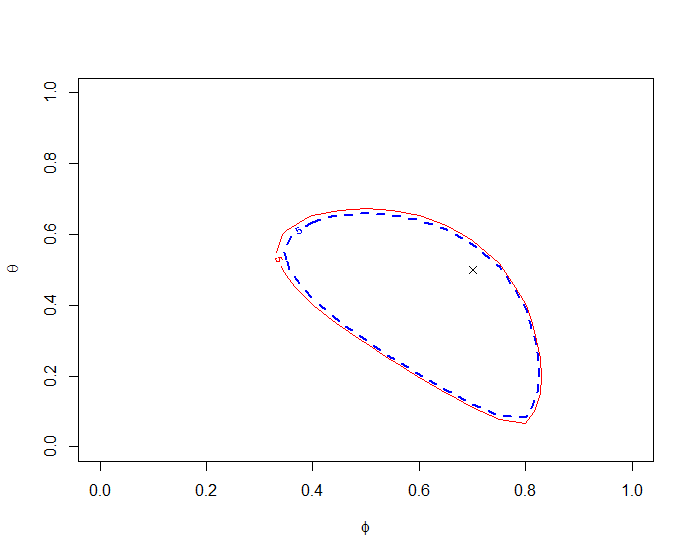}
        \caption{\small n=70}
        \label{fig:n=70}
    \end{subfigure}
    \begin{subfigure}[b]{0.3\textwidth}
        \includegraphics[width=\textwidth]{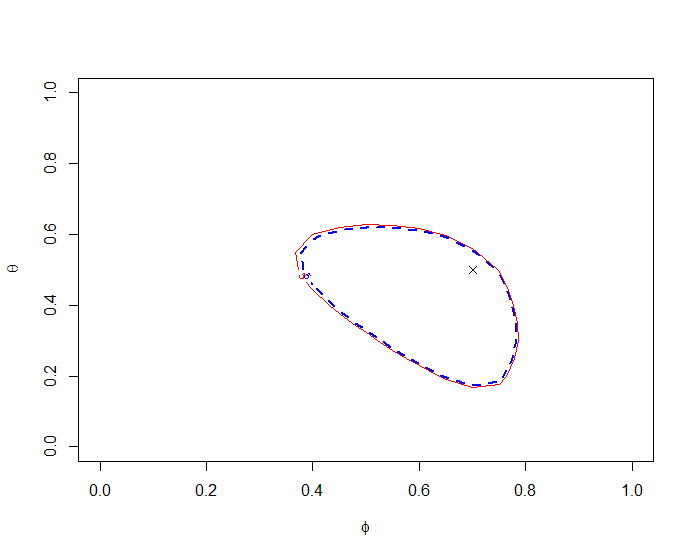}
        \caption{\small n=100}
        \label{fig:n=100}
    \end{subfigure}
    \captionsetup{justification=centering}
    \caption{\small 90\% adjusted empirical likelihood confidence region for \textsc{arma}(1,1) model with $a_t \sim N(0,1)$, `+' is the true parameter value.}\label{fig:contours}
\end{figure}

\section{Simulations: Comparison of Coverage errors of confidence sets for the parameters of \textsc{ma} and \textsc{ar} models}
\subsection{\textsc{ma(1) model}}
\noindent A Monte Carlo experiment is conducted to explore the accuracy of the adjusted empirical likelihood confidence regions for \textsc{ma} model. To make a comparison to the empirical likelihood confidence regions based on Monti's method, we choose \textsc{ma}(1), $Z_t=a_t - \theta a_{t-1}$ as Monti used. The simulations are carried out under different distributions for the white noise process $a_t$:~$N(0,1)$~and ${\chi}^2_5$ distribution centered around zero. The simulations are conducted for different values of $\theta=(0.25, 0.5, 0.7, 0.75, 0.8, 0.85, 0.9)$. In each case, 1000 series of size 20, 30, 40 and 70, are drawn and the coverage probabilities are computed. We let $a_n=\log(n)/2$ as in the definition of $\psi_{n+1}$.

The adjusted empirical likelihood coverage probabilities are compared with the unadjusted empirical likelihood coverage probabilities. Further, the coverage probabilities of intervals based on theoretical Bartlett-correction and estimated Bartlett-correction (DiCiccio et al.,1991) are used for comparison.  Table \ref{table:Table1} provides the results for the nominal level of 90\%. The values of $\theta$ in (0,1) are considered as the behavior of the coverage probabilities is almost symmetric in $\theta$. It is clear that the the coverage probabilities of the AEL are closer to the nominal value of 0.90 under each sample size and population distribution considered. Especially when the sample size is small, the AEL method gives more accurate results as compared to other methods. We also find that EL with theoretical or estimated Bartlett-correction doesn't seem to give better results than AEL.

\subsection{\textsc{ar(1) model}}
\noindent We simulate the coverage probabilities for the parameters of \textsc{ar}(1), $Z_t=\phi_1Z_{t-1} + a_t$. We consider~$N(0,1)$~and $\chi^2_5$ distribution centered around zero for the distribution of the white noise process $a_t$. We conduct 1000 simulations under the sample sizes of 20, 30, 40, 70 and $\phi$= 0.25, 0.5, 0.7, 0.9 are used. The coverage probabilities are recorded in Table 2. The coverage probabilities are substantially improved with the AEL method as compared to the EL method and EL with (estimated and theoretical) Bartlett-correction methods. The simulation results show that the AEL method gives better results than the other methods, especially with small sample sizes.

\begin{table}[ht]
\captionsetup{justification=centering}
\caption{\textit{Coverage probabilities for the parameter of} MA(1) \textit{models}} 
\centering
\small
\begin{tabular}{c c c c c c c c c}
\hline\hline 
n & Method & $\theta=0.25$ & $\theta=0.5$ & $\theta=0.7$ & $\theta=0.75$ & $\theta=0.8$ & $\theta=0.85$ & $\theta=0.9$\\ [0.5ex]
\hline
\multicolumn{9}{ c }{Model: $a_t\sim N(0,1)$} \\ [0.5ex]

\multirow{4}{*}{n=20}  & EL & 0.794 & 0.732 & 0.639 & 0.576 & 0.522 & 0.389 & 0.100 \\
 & EB & 0.807 & 0.742 & 0.655 & 0.591 & 0.535 & 0.401 & 0.103 \\
 & TB & 0.817 & 0.754 & 0.671 & 0.605 & 0.545 & 0.419 & 0.116 \\
 & AEL & 0.838 & 0.764 & 0.678 & 0.613 & 0.557 & 0.415 & 0.110 \\ [0.5ex]
\multirow{4}{*}{n=30} & EL & 0.861 & 0.817 & 0.749 & 0.717 & 0.663 & 0.624 & 0.479 \\
 & EB & 0.870 & 0.830 & 0.760 & 0.730 & 0.678 & 0.629 & 0.487 \\
 & TB & 0.874 & 0.839 & 0.770 & 0.739 & 0.689 & 0.638 & 0.498 \\
 & AEL & 0.885 & 0.847 & 0.780 & 0.749 & 0.699 & 0.634 & 0.496 \\ [0.5ex]
\multirow{4}{*}{n=40} & EL & 0.868 & 0.839 & 0.779 & 0.767 & 0.746 & 0.697 & 0.638 \\
 & EB & 0.875 & 0.849 & 0.793 & 0.770 & 0.755 & 0.711 & 0.634 \\
 & TB & 0.880 & 0.852 & 0.803 & 0.784 & 0.764 & 0.720 & 0.645 \\
 & AEL & 0.889 & 0.852 & 0.814 & 0.784 & 0.767 & 0.732 & 0.648 \\ [0.5ex]
\multirow{4}{*}{n=70} & EL & 0.880 & 0.850 & 0.831 & 0.811 & 0.777 & 0.739 & 0.694 \\
 & EB & 0.888 & 0.856 & 0.856 & 0.818 & 0.782 & 0.747 & 0.703 \\
 & TB & 0.890 & 0.861 & 0.861 & 0.837 & 0.786 & 0.751 & 0.708 \\
 & AEL & 0.894 & 0.863 & 0.863 & 0.843 & 0.790 & 0.754 & 0.715 \\ [1ex]

\multicolumn{9}{ c }{Model: $a_t\sim {\chi}^2_5 - 5$} \\ [0.5ex]
\multirow{4}{*}{n=20} & EL & 0.805 & 0.757 & 0.683 & 0.625 & 0.561 & 0.415 & 0.113 \\
 & EB & 0.817 & 0.769 & 0.694 & 0.642 & 0.570 & 0.428 & 0.118 \\
 & TB & 0.822 & 0.778 & 0.704 & 0.652 & 0.584 & 0.443 & 0.134 \\
 & AEL & 0.849 & 0.801 & 0.718 & 0.670 & 0.589 & 0.450 & 0.125 \\ [0.5ex]
\multirow{4}{*}{n=30} & EL & 0.815 & 0.789 & 0.731 & 0.702 & 0.682 & 0.635 & 0.471 \\
 & EB & 0.827 & 0.797 & 0.742 & 0.722 & 0.690 & 0.644 & 0.482 \\
 & TB & 0.836 & 0.802 & 0.750 & 0.733 & 0.703 & 0.652 & 0.493 \\
 & AEL & 0.850 & 0.811 & 0.765 & 0.742 & 0.711 & 0.661 & 0.494 \\ [0.5ex]
\multirow{4}{*}{n=40}  & EL & 0.853 & 0.816 & 0.768 & 0.748 & 0.719 & 0.676 & 0.603 \\
 & EB & 0.858 & 0.823 & 0.775 & 0.754 & 0.730 & 0.694 & 0.613 \\
 & TB & 0.862 & 0.827 & 0.782 & 0.765 & 0.742 & 0.703 & 0.622 \\
 & AEL & 0.875 & 0.829 & 0.787 & 0.766 & 0.745 & 0.708 & 0.628 \\ [0.5ex]
\multirow{4}{*}{n=70}& EL & 0.864 & 0.851 & 0.826 & 0.801 & 0.766 & 0.719 & 0.656 \\
 & EB & 0.870 & 0.855 & 0.836 & 0.836 & 0.774 & 0.727 & 0.665 \\
 & TB & 0.873 & 0.859 & 0.838 & 0.838 & 0.777 & 0.729 & 0.669 \\
 & AEL & 0.879 & 0.864 & 0.844 & 0.844 & 0.779 & 0.736 & 0.683 \\ \hline
\multicolumn{8}{ l }{EL= empirical likelihood; TB=EL with theoretical Bartlett correction;} \\
\multicolumn{8}{ l }{EB=EL with estimated Bartlett correction; AEL=Adjusted EL.} \\
\end{tabular}
\label{table:Table1}
\end{table}

\begin{table}[ht]
\captionsetup{justification=centering}
\caption{\textit{Coverage probabilities for the parameter of} AR(1) \textit{models}} 
\centering
\small
\begin{tabular}{c c c c c c}
\hline\hline 
n & Method & $\phi=0.25$ & $\phi=0.5$ & $\phi=0.7$ & $\phi=0.9$ \\ [0.5ex]
\hline
\multicolumn{6}{ c }{Model: $a_t\sim N(0,1)$} \\ [0.5ex]
\multirow{4}{*}{n=20} & EL & 0.860 & 0.799 & 0.768 & 0.476 \\
& EB & 0.869 & 0.913 & 0.779 & 0.496 \\
& TB & 0.875 & 0.822 & 0.796 & 0.552\\
& AEL & 0.892 & 0.831 & 0.807 & 0.505\\ [0.5ex]
\multirow{4}{*}{n=30} & EL & 0.870 & 0.833 & 0.800 & 0.627 \\
& EB & 0.878 & 0.844 & 0.807 & 0.651 \\
& TB & 0.890 & 0.853 & 0.816 & 0.692\\
& AEL & 0.898 & 0.860 & 0.817 & 0.656\\ [0.5ex]
\multirow{4}{*}{n=40} & EL & 0.856 & 0.859 & 0.819 & 0.683 \\
& EB & 0.867 & 0.867 & 0.831 & 0.696 \\
& TB & 0.871 & 0.871 & 0.843 & 0.742\\
& AEL & 0.878 & 0.874 & 0.838 & 0.698\\ [0.5ex]
\multirow{4}{*}{n=70} & EL & 0.860 & 0.850 & 0.849 & 0.752 \\
& EB & 0.867 & 0.862 & 0.859 & 0.761 \\
& TB & 0.870 & 0.868 & 0.868 & 0.786\\
& AEL & 0.876 & 0.870 & 0.864 & 0.763\\ [1ex]

\multicolumn{6}{ c }{Model: $a_t\sim {\chi}^2_5 - 5$} \\ [0.5ex]
\multirow{4}{*}{n=20} & EL & 0.834 & 0.788 & 0.743 & 0.505 \\
& EB & 0.845 & 0.802 & 0.757 & 0.520 \\
& TB & 0.861 & 0.821 & 0.782 & 0.578\\
& AEL & 0.877 & 0.836 & 0.780 & 0.525\\ [0.5ex]
\multirow{4}{*}{n=30} & EL & 0.840 & 0.833 & 0.790 & 0.609 \\
& EB & 0.853 & 0.843 & 0.802 & 0.629 \\
& TB & 0.862 & 0.853 & 0.821 & 0.665\\
& AEL & 0.871 & 0.856 & 0.812 & 0.634\\ [0.5ex]
\multirow{4}{*}{n=40} & EL & 0.868 & 0.837 & 0.821 & 0.695  \\
& EB & 0.871 & 0.847 & 0.826 & 0.708 \\
& TB & 0.879 & 0.857 & 0.835 & 0.749\\
& AEL & 0.883 & 0.857 & 0.831 & 0.709\\ [0.5ex]
\multirow{4}{*}{n=70} & EL & 0.870 & 0.879 & 0.826 & 0.744 \\
& EB & 0.880 & 0.886 & 0.838 & 0.765 \\
& TB & 0.881 & 0.890 & 0.849 & 0.786\\
& AEL & 0.889 & 0.890 & 0.844 & 0.764\\ [0.5ex]
\hline
\multicolumn{6}{ l }{EL= empirical likelihood; TB=EL with theoretical Bartlett correction;} \\
\multicolumn{6}{ l }{EB=EL with estimated Bartlett correction; AEL=Adjusted EL.} \\
\end{tabular}
\label{table:Table2}
\end{table}

\section{Discussion}
\noindent In this paper, we propose an adjusted empirical likelihood (AEL) to extend Monti's (1997) method for dependent observations by adopting Chen's
(2008) idea. We establish the asymptotic null distribution of the AEL statistic for stationary time series as a standard chi-square distribution which is same as the one obtained by Monti (1997). Simulations for \textsc{ar(1)}, \textsc{ma(1)} and \textsc{arma(1,1)} models with different sample sizes have been conducted to illustrate the performance of the proposed AEL method. Coverage probabilities of the AEL method have also been compared to the unadjusted EL method and the Bartlett-corrected ones. Comparisons indicate that the proposed AEL method compares favorably with the other methods, especially for the small sample sizes. Such a method is applied to a real data to construct the confidence region for the parameters of \textsc{arma} model. We observe that the confidence region obtained by the AEL contains the one obtained by the unadjusted EL while retaining the data-driven shape.

In this paper, we develop the AEL method for short-memory time series models to improve Monti's (1997) results. In our future work, we will extend this idea to different types of time series models such as long-memory time series model and change point problems for time series models.

\newpage
\appendix

\section*{Appendix A}
\begin{flushleft}
\textit{Proof of $\hat{W}^*(\beta) \sim {\chi}^2_k$}.
\end{flushleft}
\noindent First we prove that $\lambda=O_p(n^{-\frac{1}{2}})$. The adjusted empirical likelihood ratio function is,
$$W^*(\theta) = -2\,\sup \bigg \{ \sum_{j=1}^{n+1}log[(n+1)p_j]| p_j \geq 0,j=1,...n+1;\sum_{j=1}^{n+1}p_j=1; \sum_{j=1}^{n+1}\psi _j(I(\omega _j),\beta )=0 \bigg\} $$
where $\psi_{n+1} = -\frac{a_n}{n} \sum_{j=1}^{n}\psi_j = -a_n {\bar{\psi}}_n$ and $a_n=\max(1,log(n)/2)$. We will show that $W^*(\theta) \sim {\chi_k}^2$. First we need to show that $\lambda=O_p(n^{-\frac{1}{2}})$. Denote $\psi_j \equiv  \psi_j(I(\omega_j),\beta)$.\\

\noindent Assume $Var\{\psi(I(\omega),\beta)\}$ is finite and has rank $q<m (=dim(\psi))$. Let the eigenvalues of $Var\{\psi(I(\omega),\beta)\}$ be ${\sigma_1}^2,{\sigma_2}^2,...,{\sigma_m}^2$.
WLOG, assume ${\sigma_1}^2 = 1 $. Let $\lambda$ be the solution of
\begin{equation}
\sum_{j=1}^{n+1} \frac{\psi_j}{1+\lambda^\prime \psi_j} = 0. \tag{A.1}\label{eq:A1}
\end{equation}
Now let $\psi^* = \smash{\displaystyle\max_{1 \leq j \leq n}} \parallel \psi_j\parallel$. Since $\left| \psi_j \right| \geq0$ are independent, by Lemma 3 of Owen (1990), we have
$$\psi^*= \smash{\displaystyle\max_{1 \leq j \leq n}} \parallel \psi_j\parallel = o_p(n^{\frac{1}{2}}), \qquad \text{if} \qquad  E({\left|\psi_j\right|}^2)<\infty.$$
\\Thus, by CLT $\psi^*=o_p(n^{-\frac{1}{2}})$ and ${\bar{\psi}}_n=\frac{1}{n}\sum_{j=1}^{n}\psi_j=O_p(n^{-\frac{1}{2}})$. Also, $\lambda=o_p(1)$.
\\Let $\lambda=\rho\theta$ where $\rho\geq0$ and $\parallel\theta\parallel=1$. Multiplying both sides of \eqref{eq:A1} by $n^{-1}\theta^\prime$, we get
\begin{alignat*}{1}
0 &= n^{-1}\theta^\prime \sum_{j=1}^{n+1}\frac{\psi_j}{1+\lambda^\prime\psi_j} \\
&= \frac{\theta^\prime}{n}\sum_{j=1}^{n+1}\bigg[ \psi_j - \frac{\lambda^\prime{\psi_j}^2}{1+\lambda^\prime\psi_j}\bigg]\\
&= \frac{\theta^\prime}{n}\sum_{j=1}^{n+1} \psi_j - \frac{\rho}{n} \sum_{j=1}^{n+1}\frac{({\theta^\prime\psi_j})^2}{1+\rho\theta^\prime\psi_j}. \tag{A.2}\label{eq:A2}
\end{alignat*}
Next consider the first term in \eqref{eq:A2},
\begin{alignat*}{1}
\frac{\theta^\prime}{n}\sum_{j=1}^{n+1} \psi_j &= \frac{\theta^\prime}{n}\sum_{j=1}^{n} \psi_j + \frac{\theta^\prime}{n} \psi_{n+1} \\
&= \theta^{\prime} \bigg[ \frac{1}{n} \sum_{j=1}^{n}\psi_j + \frac{1}{n} \psi_{n+1} \bigg]\\
&= \theta^{\prime} \bigg[ \bar{\psi}_n + \frac{1}{n} (-a_n\bar{\psi}_n) \bigg]\\
&= \theta^{\prime} \bar{\psi}_n \bigg( 1-\frac{a_n}{n} \bigg).
\end{alignat*}
Consider the second term in \eqref{eq:A2},
\begin{alignat*}{1}
\frac{\rho}{n} \sum_{j=1}^{n+1}\frac{({\theta^\prime\psi_j})^2}{1+\rho\theta^\prime\psi_j}
&= \frac{\rho}{n} \sum_{j=1}^{n+1} ({\theta^\prime\psi_j})^2 \frac{1}{1+\rho\theta^\prime\psi_j} \tag{$\parallel\rho\theta^{\prime}\psi_j\parallel = \rho\parallel\psi_j\parallel$ since $\parallel\theta\parallel=1$}\\
&\geq \frac{\rho}{n} \sum_{j=1}^{n+1} \frac{({\theta^\prime\psi_j})^2}{1+\rho\theta^\prime\psi^*}.
\end{alignat*}
So we have,
\begin{alignat*}{1}
0&= n^{-1}\theta^\prime \sum_{j=1}^{n+1}\frac{\psi_j}{1+\lambda^\prime\psi_j}\\
&= \frac{\theta^\prime}{n}\sum_{j=1}^{n+1}\psi_j - \frac{\rho}{n} \sum_{j=1}^{n+1}\frac{({\theta^\prime\psi_j})^2}{1+\rho\theta^\prime\psi_j}\\
&\leq \theta^\prime \bar{\psi}_n \bigg( 1-\frac{a_n}{n}\bigg) - \frac{\rho}{n(1+\rho\psi^*)} \sum_{j=1}^{n}(\theta^\prime\psi_j)^2 \tag{$(n+1)^{th}$ term in the second summation is non-negative}\\
&= \theta^\prime \bar{\psi}_n - \theta^\prime {\bar{\psi}}_n \frac{a_n}{n} - \frac{\rho}{n(1+\rho\psi^*)}\sum_{j=1}^{n}(\theta^\prime\psi_j)^2\\
&=\theta^\prime \bar{\psi}_n - \frac{\rho}{n(1+\rho\psi^*)}\sum_{j=1}^{n}(\theta^\prime\psi_j)^2 + O_p(n^{-\frac{3}{2}}a_n). \tag{A.3} \label{eq:A3}
\end{alignat*}
The assumption on $Var\{\psi(I(\omega),\beta)\}$ implies that
\begin{align*}
\frac{1}{n}\sum_{j=1}^{n+1}(\theta^\prime\psi_j)^2 \geq (1-\epsilon){\sigma_1}^2 = 1-\epsilon
\end{align*}
in probability for some $0 < \epsilon <1$. So as long as $a_n = o_p(n)$, \eqref{eq:A3} implies that
\begin{align*}
\frac{\rho}{1+\rho\psi^*} &\leq \theta^\prime\bar{\phi}_n(1-\epsilon)^{-1} = o_p(n^{-\frac{1}{2}}).
\end{align*}
Since
$$\theta^\prime \bar{\psi}_n - \frac{\rho}{n(1+\rho\psi^*)}\sum_{j=1}^{n}(\theta^\prime\psi_j)^2 \geq 0, $$
$$\theta^\prime\bar{\psi}_n \geq \frac{\rho}{(1+\rho\psi^*)}\frac{\sum_{j=1}^{n}(\theta^\prime\psi_j)^2}{n} \geq \frac{\rho}{(1+\rho\psi^*)} (1-\epsilon). $$
\noindent Hence $$\frac{\rho}{(1+\rho\psi^*)} \leq \theta^\prime \bar{\psi}_n (1-\epsilon)^{-1}.$$
Therefore, $\rho = \parallel\lambda\parallel=O_p(n^{-\frac{1}{2}})$ which implies $\lambda=O_p(n^{-\frac{1}{2}})$.\\

Now we need to prove that $W^*(\theta) \sim {\chi_k}^2$. Let $\beta=\beta_{n+1} + n^{-\frac{1}{2}}u$ with $\left| u \right| < +\infty$.
\\Using Taylor series expansion on \eqref{eq:A1},
\begin{align*}
\frac{1}{n+1}\sum_{j=1}^{n+1}\psi_j [1 - \lambda^\prime \psi_j] = O(n^{-1}),
\end{align*}
\begin{equation}
\frac{1}{n+1}\sum_{j=1}^{n+1}\psi_j  - \frac{1}{n+1}\sum_{j=1}^{n+1}\psi_j \lambda^\prime {\psi_j}^\prime = O(n^{-1}). \tag{A.4} \label{eq:A4}
\end{equation}
So,
\begin{equation}
\frac{1}{n+1}\sum_{j=1}^{n+1}\psi_j = \frac{1}{n+1}\sum_{j=1}^{n+1}\frac{\partial \psi_j(I(\omega_j),t)}{\partial t^\prime} \bigg|_{\scriptscriptstyle \beta_{n+1}} (\beta-\beta_{n+1}) + O(n^{-1}), \tag{A.5} \label{eq:A5}
\end{equation}
and
\begin{equation}
\frac{1}{n+1}\sum_{j=1}^{n+1}\psi_j {\psi_j}^\prime = \frac{1}{n+1}\sum_{j=1}^{n+1}\psi_j(I(\omega_j),\beta_{n+1}){\psi_j(I(\omega_j),\beta_{n+1})}^\prime + O(n^{-\frac{1}{2}}). \tag{A.6} \label{eq:A6}
\end{equation}
Let
\begin{equation}
\hat{A}(\beta_{n+1}) = \frac{1}{n+1}\sum_{j=1}^{n+1}\frac{\partial \psi_j(I(\omega_j),t)}{\partial t^\prime} \bigg|_{\scriptscriptstyle \beta_{n+1}}, \tag{A.7} \label{eq:A7}
\end{equation}
and
\begin{equation}
\hat{\Sigma}(\beta_{n+1}) = \frac{1}{n+1}\sum_{j=1}^{n+1}\psi_j(I(\omega_j),\beta_{n+1}){\psi_j(I(\omega_j),\beta_{n+1})}^\prime. \tag{A.8} \label{eq:A8}
\end{equation}
Using \eqref{eq:A5} and \eqref{eq:A6} in \eqref{eq:A4} we get,
$$\hat{A}(\beta_{n+1}) (\beta-\beta_{n+1}) + O(n^{-1}) - \lambda^\prime \hat{\Sigma}(\beta_{n+1}) - O(n^{-\frac{1}{2}}) = O(n^{-1}),$$
$$\lambda^\prime \hat{\Sigma}(\beta_{n+1}) = \hat{A}(\beta_{n+1}) (\beta-\beta_{n+1}) +  O(n^{-\frac{1}{2}}),$$
\begin{equation}
\lambda = { \hat{\Sigma}(\beta_{n+1})}^{-1} \hat{A}(\beta_{n+1}) (\beta-\beta_{n+1}) +  O(n^{-1}). \tag{A.9} \label{eq:A9}
\end{equation}
Expanding $W^*(\beta) = 2 \sum_{j=1}^{n+1}\log [1+\lambda^\prime \psi_j(I(\omega_j),\beta)]$ by Taylor expansion, we obtain,
\begin{align*}
W^*(\beta) &= 2 \sum_{j=1}^{n+1}\log [1+\lambda^\prime \psi_j(I(\omega_j),\beta)]\\
&= 2\sum_{j=1}^{n+1} \big[ \lambda^\prime \psi_j(I(\omega_j),\beta) - \frac{1}{2} {(\lambda^\prime \psi_j(I(\omega_j),\beta))}^2 \big] + O(n^{-\frac{1}{2}}).
\end{align*}
and using \eqref{eq:A5}, \eqref{eq:A6} and \eqref{eq:A9} gives,
\begin{align*}
W^*(\beta) = (n+1)(\beta-\beta_{n+1})^\prime \hat{V}^{-1} (\beta-\beta_{n+1}) + O(n^{-\frac{1}{2}}),  \tag{A.10} \label{eq:A10}
\end{align*}
where
$$ \hat{V} = {\hat{A}(\beta_{n+1})}^{-1} \hat{\Sigma}(\beta_{n+1}) {\{ {\hat{A}(\beta_{n+1})}^\prime\} }^{-1}, $$
which converges to a standard chi-square distribution with $k$ degrees of freedom as $n \rightarrow \infty$.\\

\noindent As follows, we provide the details of obtaining \eqref{eq:A10}. We have
$$\lambda = {\hat{\Sigma}(\beta_{n+1})}^{-1} \hat{A}(\beta_{n+1}) (\beta-\beta_{n+1}) + O(n^{-\frac{1}{2}}),$$
and
\begin{align*}
W^*(\beta) = 2\sum_{j=1}^{n+1} \big[ \lambda^\prime\psi_j - \frac{1}{2} {(\lambda^\prime\psi_j)}^2 \big] + O(n^{-\frac{1}{2}}) \\
=  2\sum_{j=1}^{n+1}\lambda^\prime\psi_j - \sum_{j=1}^{n+1}{(\lambda^\prime\psi_j)}^2 + O(n^{-\frac{1}{2}}). \tag{R.1} \label{R.1}
\end{align*}
Here $\psi_j \equiv  \psi_j(I(\omega_j),\beta)$. We know $\bar{\psi}_n= O_p(n^{-\frac{1}{2}})$ and $\psi_{n+1}=-a_n\bar{\psi}_n$ and $a_n = o_p(n)$. Therefore $\psi_{n+1}=O(n^{\frac{1}{2}})$. So,
\begin{align*}
\frac{1}{n+1}\sum_{j=1}^{n+1}\psi_j &= \frac{n}{n+1}\bar{\psi}_n + \frac{1}{n+1}\psi_{n+1} = \frac{n}{n+1} O(n^{-\frac{1}{2}}) + \frac{1}{n+1} O(n^{\frac{1}{2}})\\
&= O(n^{-\frac{1}{2}}) + O(n^{-\frac{1}{2}}) =O(n^{-\frac{1}{2}}).
\end{align*}
From \eqref{eq:A5}
$$\frac{1}{n+1}\sum_{j=1}^{n+1}\psi_j = \hat{A}(\beta_{n+1}) (\beta-\beta_{n+1}) + O(n^{-1}), $$
$$O(n^{-\frac{1}{2}}) = \hat{A}(\beta_{n+1}) (\beta-\beta_{n+1}) + O(n^{-1}),$$
 $\hat{A}(\beta_{n+1}) (\beta-\beta_{n+1}) = O(n^{-\frac{1}{2}})$.\\
\\Now consider the first term of \eqref{R.1},
\begin{align*}
\sum_{j=1}^{n+1}\lambda^\prime\psi_j &= \lambda^\prime(n+1)\big[ \hat{A}(\beta_{n+1}) (\beta-\beta_{n+1}) +O(n^{-1})\big]\\
&= \lambda^\prime(n+1)\hat{A}(\beta_{n+1}) (\beta-\beta_{n+1}) + \lambda^\prime(n+1) O(n^{-1})\\
&= \lambda^\prime(n+1)\hat{A}(\beta_{n+1}) (\beta-\beta_{n+1}) +  O(n^{-\frac{1}{2}}).\\
\end{align*}
Since $\lambda=O(n^{-\frac{1}{2}})$,
\begin{align*}
\sum_{j=1}^{n+1}\lambda^\prime\psi_j &= (n+1) (\beta-\beta_{n+1})^\prime {\hat{A}(\beta_{n+1})}^\prime {{\hat{\Sigma}(\beta_{n+1})}^{-1}}^\prime \hat{A}(\beta_{n+1})(\beta-\beta_{n+1})\\ + &O(n^{-1})(n+1)\hat{A}(\beta_{n+1})(\beta-\beta_{n+1}) + O(n^{-\frac{1}{2}})\\
&= (n+1) (\beta-\beta_{n+1})^\prime {\hat{A}(\beta_{n+1})}^\prime {{\hat{\Sigma}(\beta_{n+1})}^{-1}}^\prime \hat{A}(\beta_{n+1})(\beta-\beta_{n+1}) + O(n^{-\frac{1}{2}}). \tag{R.2}\label{R2}
\end{align*}
\\From \eqref{eq:A6}, we have
\begin{align*}
\sum_{j=1}^{n+1}\psi_j {\psi_j}^\prime =(n+1)\hat{\Sigma}(\beta_{n+1}) + (n+1)O(n^{-\frac{1}{2}}) = (n+1)\hat{\Sigma}(\beta_{n+1}) + O(n^{\frac{1}{2}}).
\end{align*}
Therefore,
\begin{align*}
{\lambda^\prime}^2 \sum_{j=1}^{n+1}{\psi_j}^2 = {\lambda^\prime}^2 (n+1) \hat{\Sigma}(\beta_{n+1}) + {\lambda^\prime}^2 O(n^{\frac{1}{2}}) = {\lambda^\prime}^2 (n+1) \hat{\Sigma}(\beta_{n+1}) + O(n^{-\frac{1}{2}})
\end{align*}
From \eqref{eq:A9},
\begin{align*}
{\lambda^\prime}^2 &= \lambda^\prime\lambda = \big[ (\beta-\beta_{n+1})^\prime {\hat{A}(\beta_{n+1})}^\prime {({\hat{\Sigma}(\beta_{n+1})}^{-1})}^\prime  + O(n^{-1})\big] \big[{\hat{\Sigma}(\beta_{n+1})}^{-1} \hat{A}(\beta_{n+1}) (\beta-\beta_{n+1}) + O(n^{-1})\big]\\
&= (\beta-\beta_{n+1})^\prime {\hat{A}(\beta_{n+1})}^\prime {({\hat{\Sigma}(\beta_{n+1})}^{-1})}^\prime {\hat{\Sigma}(\beta_{n+1})}^{-1} \hat{A}(\beta_{n+1}) (\beta-\beta_{n+1})\\ &+(\beta-\beta_{n+1})^\prime {\hat{A}(\beta_{n+1})}^\prime {({\hat{\Sigma}(\beta_{n+1})}^{-1})}^\prime O(n^{-1})+ O(n^{-1}) {\hat{\Sigma}(\beta_{n+1})}^{-1} \hat{A}(\beta_{n+1}) (\beta-\beta_{n+1}) + O(n^{-2}) \\
&=  (\beta-\beta_{n+1})^\prime {\hat{A}(\beta_{n+1})}^\prime {({\hat{\Sigma}(\beta_{n+1})}^{-1})}^\prime {\hat{\Sigma}(\beta_{n+1})}^{-1} \hat{A}(\beta_{n+1}) (\beta-\beta_{n+1}) + 2 O(n^{-\frac{3}{2}}){({\hat{\Sigma}(\beta_{n+1})}^{-1})}^\prime + O(n^{-2}).
\end{align*}
Now consider the second term of \eqref{R.1},
\begin{align*}
&{\lambda^\prime}^2 \sum_{j=1}^{n+1}{\psi_j}^2 = {\lambda^\prime}^2 (n+1) \hat{\Sigma}(\beta_{n+1}) + O(n^{-\frac{1}{2}})\\
&= \big[ (\beta-\beta_{n+1})^\prime {\hat{A}(\beta_{n+1})}^\prime {({\hat{\Sigma}(\beta_{n+1})}^{-1})}^\prime {\hat{\Sigma}(\beta_{n+1})}^{-1} \hat{A}(\beta_{n+1}) (\beta-\beta_{n+1})+2 O(n^{-\frac{3}{2}}){({\hat{\Sigma}(\beta_{n+1})}^{-1})}^\prime + O(n^{-2}) \big] \\
&(n+1) \hat{\Sigma}(\beta_{n+1}) + O(n^{-\frac{1}{2}})\\
&= (n+1) (\beta-\beta_{n+1})^\prime {\hat{A}(\beta_{n+1})}^\prime {\hat{\Sigma}(\beta_{n+1})}^{-1} \hat{A}(\beta_{n+1}) (\beta-\beta_{n+1}) + O(n^{-\frac{1}{2}}) + O(n^{-1}) \hat{\Sigma}(\beta_{n+1}). \tag{R.3} \label{R3}
\end{align*}
Rewriting \eqref{R.1} in terms of \eqref{R2} and \eqref{R3} we obtain,
\begin{align*}
&W^*(\beta) = 2\, \eqref{R2} - \eqref{R3} +  O(n^{-\frac{1}{2}})\\
&= 2 \,  \bigg[ (n+1) (\beta-\beta_{n+1})^\prime {\hat{A}(\beta_{n+1})}^\prime {{\hat{\Sigma}(\beta_{n+1})}^{-1}}^\prime \hat{A}(\beta_{n+1})(\beta-\beta_{n+1}) + O(n^{-\frac{1}{2}}) \bigg] \\
&- \bigg[ (n+1) (\beta-\beta_{n+1})^\prime {\hat{A}(\beta_{n+1})}^\prime {\hat{\Sigma}(\beta_{n+1})}^{-1} \hat{A}(\beta_{n+1}) (\beta-\beta_{n+1}) + O(n^{-\frac{1}{2}}) + O(n^{-1}) \hat{\Sigma}(\beta_{n+1}) \bigg] + O(n^{-\frac{1}{2}})\\
&=  (n+1) (\beta-\beta_{n+1})^\prime {\hat{A}(\beta_{n+1})}^\prime {\hat{\Sigma}(\beta_{n+1})}^{-1} \hat{A}(\beta_{n+1}) (\beta-\beta_{n+1}) + O(n^{-\frac{1}{2}}) + O(n^{-1}) \hat{\Sigma}(\beta_{n+1})\\
&= (n+1) (\beta-\beta_{n+1})^\prime {\hat{A}(\beta_{n+1})}^\prime {\hat{\Sigma}(\beta_{n+1})}^{-1} \hat{A}(\beta_{n+1}) (\beta-\beta_{n+1}) + O(n^{-\frac{1}{2}}).
\end{align*}

\section*{Appendix A}
\begin{flushleft}
\textit{Proof of Asymptotic distribution of (5.1)}:
\end{flushleft}
\section*{Appendix B}
\noindent \textit{Proof of Asymptotic distribution of (5.1)}:
\\Using the same process as in Appendix A, we can show that
$$W^*(\beta_{(1)}) = (n+1) (\beta_{(1)}-\beta_{(1),n+1})^\prime \hat{V}^{-1}_1 (\beta_{(1)}-\beta_{(1),n+1}) + O(n^{-\frac{1}{2}})  $$
where $\beta_{(1),n+1}$ is the estimator,
$$ \hat{V_1} = {\hat{A}(\beta_{(1),n+1})}^{-1} \hat{\Sigma}(\beta_{(1),n+1}) {\{ {\hat{A}(\beta_{(1),n+1})}^\prime\} }^{-1},$$
$$\hat{A}(\beta_{(1),n+1}) = \frac{1}{n+1}\sum_{j=1}^{n+1}\frac{\partial {\tilde{\psi}}_j(I(\omega_j),t)}{\partial t^\prime} \bigg|_{\scriptscriptstyle \beta_{(1),n+1}},$$
and,
$$\hat{\Sigma}(\beta_{(1),n+1}) = \frac{1}{n+1}\sum_{j=1}^{n+1}{\tilde{\psi}}_j(I(\omega_j),\beta_{(1),n+1}){{\tilde{\psi}}_j(I(\omega_j),\beta_{(1),n+1})}^\prime .$$
Since,
$${\tilde{\psi}}_j(I(\omega_j),\beta_{(1),n+1} \rightarrow {\tilde{\psi}}_j(I(\omega_j),\beta_{(1)} = \frac{I(\omega_j)}{g^1_j(\beta_{(1)})} \frac{\partial\ln\{g^1_j(t)\}}{\partial t} \bigg|_{\scriptscriptstyle \beta_{(1)}},$$
\begin{align*}
&\frac{\partial {\tilde{\psi}}_j\{I(\omega_j),t\}}{\partial t^\prime}\bigg|_{\scriptscriptstyle \beta_{(1),n+1}} \rightarrow \frac{\partial {\tilde{\psi}}_j\{I(\omega_j),t\}}{\partial t^\prime}\bigg|_{\scriptscriptstyle \beta_{(1)}}\\
&= \frac{I(\omega_j)}{g^1_j(\beta_{(1)})} \Bigg[ \frac{\partial^2\ln\{g^1_j(t)\}}{\partial t \partial t^\prime} \bigg|_{\scriptscriptstyle \beta_{(1)}} - \frac{\partial\ln\{g^1_j(t)\}}{\partial t} \bigg|_{\scriptscriptstyle \beta_{(1)}} \frac{\partial\ln\{g^1_j(t)\}}{\partial t^\prime} \bigg|_{\scriptscriptstyle \beta_{(1)}} \Bigg]
\end{align*}
in probability, $\hat{A}(\beta_{(1),n+1})$ and  $\hat{\Sigma}(\beta_{(1),n+1})$ have the same asymptotic behavior as
$$\hat{A}(\beta_{(1)}) = \frac{1}{n+1}\sum_{j=1}^{n+1}\frac{\partial {\tilde{\psi}}_j(I(\omega_j),t)}{\partial t^\prime} \bigg|_{\scriptscriptstyle \beta_{(1)}}, \qquad \hat{\Sigma}(\beta_{(1)}) = \frac{1}{n+1}\sum_{j=1}^{n+1}{\tilde{\psi}}_j(I(\omega_j),\beta_{(1)}){{\tilde{\psi}}_j(I(\omega_j),\beta_{(1)})}^\prime .$$


\begin{thebibliography}{99}
\expandafter\ifx\csname natexlab\endcsname\relax\def\natexlab#1{#1}\fi

\bibitem[{Box \& Jenkins(1976)}]{Box:1976}
\textsc{Box, G.~E.~P.} \& \textsc{Jenkins, G.~M.} (1976).
\newblock \textit{{Time Series Analysis: Forecasting \& Control.}}
\newblock {San Francisco:}
\newblock {Holden Day.}

\bibitem[{Chan et al.(2006)Chan and Ling}]{Chan:Ling:2006}
\textsc{Chan, N.H.} and \textsc{Ling, S.}
  (2006).
\newblock Empirical Likelihood for GARCH models.
\newblock \textit{Econometric Theory} \textbf{22}, 402-428.

\bibitem [{Chan et al.(2006)Chan and Liu}]{Chan:Liu:2010}
\textsc{Chan, N.H.} and \textsc{Liu, L.} {2010.}
\newblock Bartlett Correctability of Empirical Likelihood in Time Series. 
\newblock \textit{Journal of The Japanese Statistical Society} \textbf{40}, 1-5.

\bibitem[{Chen et~al.(2006)Chen, Variyath \&
  Abraham}]{Chen:Variyath:Abraham:2008}
\textsc{Chen, J.}, \textsc{Variyath, A.~M.} \& \textsc{Abraham, B.}
  (2008).
\newblock Adjusted Empirical Likelihood and its Properties.
\newblock \textit{Journal of Computational and Graphical Statistics} \textbf{17}, 426-443.



\bibitem[{Diciccio.(1989) Diciccio, Hall \&
  Romano }]{Diciccio:Hall: Romano:1991}
\textsc{Diciccio, T.~J.}, \textsc{Hall, P.} \& \textsc{ Romano, J.~P.}
  (1991).
\newblock Empirical Likelihood is Bartlett-correctable.
\newblock \textit{The Annals of Statistics} \textbf{19}, 1053-61.


\bibitem[{Kitamura (1997)}]{Kitamura:1997}
\textsc{Kitamura, Y.} (1997).
\newblock Empirical Likelihood Methods with Weakly Dependent Processes. 
\newblock \textit{The Annals of Statistics} \textbf{25}, 2084-2102.


\bibitem[{Monti (1997)}]{Monti:1997}
\textsc{Monti, A.~C.} (1997).
\newblock Empirical Likelihood Confidence Regions in Time Series Models.
\newblock \textit{Biometrika} \textbf{84}, 395-405.



\bibitem[{Mykland (1995)}]{Mykland:1995}
\textsc{Mykland, P.A.} (1995).
\newblock Dual Likelihood. 
\newblock \textit{The Annals of Statistics} \textbf{23}, 396-421.


\bibitem[{Owen (1988)}]{Owen:1988}
\textsc{Owen, A.~B.} (1988).
\newblock Empirical Likelihood Ratio Confidence Intervals for a Single Functional.
\newblock \textit{Biometrika} \textbf{75}, 237-49.

\bibitem[{Owen (1990)}]{Owen:1990}
\textsc{Owen, A.~B.} (1990).
\newblock Empirical Likelihood Ratio Confidence Regions.
\newblock \textit{ The Annals of Statistics } \textbf{18}, 90-120.

\bibitem[{Owen (1991)}]{Owen:1991}
\textsc{Owen, A.~B.} (1991).
\newblock Empirical Likelihood for Linear Models.
\newblock \textit{The Annals of Statistics } \textbf{19}, 1725-47.

\bibitem[{Owen (2001)}]{Owen:2001}
\textsc{Owen, A.~B.} (2001).
\newblock \textit{{Empirical Likelihood.}}
\newblock {New York:}
\newblock {Chapman \& Hall/CRC.}

\bibitem[{Whittle (1953)}]{Whittle:1953}
\textsc{Whittle, P.} (1953).
\newblock Estimation and Information in Time Series.
\newblock \textit{Arkiv for Matematik} \textbf{2}, 423-34.


\end{thebibliography}

\end{document}